
\input lanlmac

\input epsf

\newcount\figno
\figno=0
\def\fig#1#2#3{
\par\begingroup\parindent=0pt\leftskip=1cm\rightskip=1cm\parindent=0pt
\baselineskip=11pt
\global\advance\figno by 1
\midinsert
\epsfxsize=#3
\centerline{\epsfbox{#2}}
\vskip 12pt
\centerline{{\bf Fig. \the\figno:} #1}\par
\endinsert\endgroup\par
}
\def\figlabel#1{\xdef#1{\the\figno}}


\def\O{{\cal O}}

\def\th{\theta}

\def\cob{\delta}
\def\ep{\epsilon}

\def\S{{\bf S}}

\def\Tr{{\rm Tr}}

\def\hf{{1\over 2}}

\def\R{{\bf R}}
\def\o{\over}

\def\til#1{\widetilde{#1}}

\def\si{\sigma}
\def\Si{\Sigma}

\def\del{\partial}

\def\bra{\langle}
\def\ket{\rangle}
\def\lf{\left}
\def\ri{\right}
\def\riya{\rightarrow}

\def\h#1{\widehat{#1}}

\def\st{\star}

\def\P{{\bf P}}

\def\sitarel#1#2{\mathrel{\mathop{\kern0pt #1}\limits_{#2}}}

\def\np#1#2#3{{ Nucl. Phys.} {\bf B#1} (#2) #3}

\def\pln#1#2#3{{Phys. Lett.} {\bf B#1} (#2) #3}

\def\hpt#1{{\tt hep-th/#1}}

\lref\SW{
N. Seiberg and E. Witten,
``String Theory and Noncommutative Geometry'',
JHEP {\bf 9909} (1999) 032, \hpt{9908142}.
}

\lref\OkawaO{
Y. Okawa and H. Ooguri,
``How Noncommutative Gauge Theories Couple to Gravity'',
\hpt{0012218}.
}
\lref\LiuM{
H. Liu and J. Michelson, 
``Supergravity Couplings of Noncommutative D-branes'',
\hpt{0101016}.
}
\lref\ReyU{
S.-J. Rey and R. Unge,
``S-Duality, Noncritical Open String and Noncommutative Gauge Theory'',
\hpt{0007089}.
}
\lref\DasR{
S. Das and S.-J. Rey,
``Open Wilson Lines in Noncommutative Gauge Theory 
and Tomography of Holographic Dual Supergravity'',
\np{590}{2000}{453}, \hpt{0008042}.
}
\lref\IIKK{
N. Ishibashi, S. Iso, H. Kawai and Y. Kitazawa,
``Wilson Loops in Noncommutative Yang Mills'',
\np{573}{2000}{573}, \hpt{9910004}.
}
\lref\AmbjornMNS{
J. Ambjorn, Y. M. Makeenko, J. Nishimura and R. J. Szabo,
``Finite N Matrix Models of Noncommutative Gauge Theory'',
JHEP {\bf 9911} (1999) 029, \hpt{9911041};
``Nonperturbative Dynamics of Noncommutative Gauge Theory'',
\pln{480}{2000}{399}, \hpt{0002158};
``Lattice Gauge Fields and Discrete Noncommutative Yang-Mills Theory'',
JHEP {\bf 0005} (2000) 023, \hpt{0004147}.
}

\lref\GrossHI{
D. J. Gross, A. Hashimoto and N. Itzhaki,
``Observables of Non-Commutative Gauge Theories'',
\hpt{0008075}.
}
\lref\DharW{
A. Dhar and S. R. Wadia, 
``A Note on Gauge Invariant Operators in Noncommutative 
Gauge Theories and the Matrix Model'',
\pln{495}{2000}{413}, \hpt{0008144}.
}
\lref\DharK{
A. Dhar and Y. Kitazawa,
``Wilson Loops in Strongly Coupled Noncommutative Gauge Theories'',
\hpt{0010256}.
}
\lref\DharKtwo{
A. Dhar and Y. Kitazawa,
``High Energy Behavior of Wilson Lines'',
\hpt{0012170}.
}
\lref\RozaliR{
M. Rozali and M. V. Raamsdonk, 
``Gauge Invariant Correlators in Non-Commutative Gauge Theory'',
\hpt{0012065}.
}

\lref\Garousi{
M. R. Garousi,
``Non-commutative world-volume interactions on D-brane 
and Dirac-Born-Infeld action'',
\np{579}{2000}{209}, \hpt{9909214}.
}
\lref\startrekI{
H. Liu and J. Michelson, 
``$*$-TREK: The One-Loop N=4 Noncommutative SYM Action'',
\hpt{0008205}.
}
\lref\MehenW{
T. Mehen and M. B. Wise,
``Generalized $*$-Products, Wilson Lines and 
the Solution of the Seiberg-Witten Equations'',
JHEP {\bf 0012} (2000) 008, \hpt{0010204}.
}
\lref\startrekII{
H. Liu, 
``$*$-Trek II: $*_n$ Operations, Open Wilson Lines 
and the Seiberg-Witten Map'',
\hpt{0011125}.
}
\lref\DasT{
S. R. Das and S. P. Trivedi,
``Supergravity couplings to Noncommutative Branes, 
Open Wilson Lines and Generalised Star Products'',
\hpt{0011131}.
}
\lref\PerniciSZ{
M. Pernici, A. Santambrogio and D. Zanon,
``The one-loop effective action of noncommutative 
${\cal N}=4$ super Yang-Mills is gauge invariant'',
\hpt{0011140}.
}
\lref\GarousiII{
M. R. Garousi,
``Transformation of the Dirac-Born-Infeld action under 
the Seiberg-Witten map'', \hpt{0011147}.
}
\lref\KiemPL{
Y. Kiem, D. H. Park and S. Lee,
``Factorization and generalized $*$-products'',
\hpt{0011233}.
}
\lref\KumarMP{
A. Kumar, A. Misra and K. L. Panigrahi,
``Noncommutative N=2 Strings'',
\hpt{0011206}.
}

\lref\Okuyama{
K. Okuyama,
``A Path Integral Representation of the Map between 
Commutative and Noncommutative Gauge Fields'',
JHEP {\bf 0003} (2000) 016, \hpt{9910138}.
}
\lref\Ishibash{
N. Ishibashi,
``A Relation between Commutative and Noncommutative 
Descriptions of D-branes'',
\hpt{9909176}.
}
\lref\Kont{
M. Kontsevich,
``Deformation quantization of Poisson manifolds'',
{\tt q-alg/9709040}.
}
\lref\CF{
A. S. Cattaneo and G. Felder,
``A path integral approach to the Kontsevich quantization formula'',
{\tt math.QA/9902090}.
}
\lref\JurcoS{
B. Jurco and P. Schupp,
``Noncommutative Yang-Mills from equivalence of star products'',
Eur. Phys. J. {\bf C14} (2000) 367, \hpt{0001032}\semi
B. Jurco, P. Schupp and J. Wess,
``Noncommutative gauge theory for Poisson manifolds'',
Nucl. Phys. {\bf B584} (2000) 784, \hpt{0005005};
``Nonabelian noncommutative gauge fields and Seiberg-Witten map'',
\hpt{0012225}.
}
\lref\Rudychev{
I. Rudychev,
``From noncommutative string/membrane to ordinary ones'',
\hpt{0101039}.
}

\Title{             
                                             \vbox{\hbox{KEK-TH-740}
                                             \hbox{hep-th/0101177}}}
{\vbox{
\centerline{Comments on Open Wilson Lines }
\vskip 4mm
\centerline{and Generalized Star Products} 
}}

\vskip .2in

\centerline{Kazumi Okuyama}

\vskip .2in

\centerline{\sl Theory Group, KEK,  Tsukuba, Ibaraki 305-0801, Japan}
\centerline{\tt kazumi@post.kek.jp}

\vskip 3cm
\noindent

We consider an open Wilson line as a momentum representation of
a boundary state which describes a $D$-brane in a constant $B$-field
background.  
Using this picture, we study the Seiberg-Witten map 
which relates the commutative and noncommutative gauge fields, 
and determine the products of fields appearing in the
general terms in the expansion of this map.

\Date{January 2001}

\vfill
\vfill

\newsec{Introduction}
In the recent study of noncommutative gauge theories,
the open Wilson lines 
play an important role.
They are used to construct gauge invariant operators 
\refs{\IIKK,\AmbjornMNS,\GrossHI,\DharW},
and also appear in the couplings
between open and closed string modes \refs{\ReyU,\DasR,\DasT,\LiuM,\OkawaO}.
The correlation functions of open Wilson lines are studied in 
\refs{\GrossHI,\DharK,\RozaliR,\DharKtwo}.

In the expansion of an open Wilson line as a  power series of the 
noncommutative gauge field, there appears the so-called generalized star 
products \startrekII. 
These products also appear in the scattering amplitudes 
of string theories and noncommutative field theories
\refs{\Garousi,\startrekI,\PerniciSZ,\GarousiII,\KumarMP,\KiemPL}.
In \refs{\MehenW,\startrekII}, open Wilson lines were 
used to construct the Seiberg-Witten map which relates
the commutative and noncommutative gauge fields \SW.
In this map, there also appears the structure of the generalized
star products. 

In this short note, we consider an open Wilson line as a
momentum representation of a boundary state describing a $D$-brane
in a constant $B$-field background. Using this picture,
we study the Seiberg-Witten map 
and determine the products of fields which appear in the general terms 
in the expansion of this map. 

This paper is organized as follows: In section 2, we review the
open Wilson lines from the viewpoint of boundary state.
In section 3, we show that the generalized star products
can be written as correlation functions on $\S^1$.
In section 4, we construct the Seiberg-Witten map
up to $\O(A^3)$ from the gauge equivalence condition and
compare this result with that obtained from the open Wilson line.
Section 5 is devoted to the conclusion and discussion. 

\newsec{Open Wilson Lines from Boundary State}
In this section, 
we review how open contours appear in  gauge invariant
objects in noncommutative gauge theories \refs{\IIKK,\DharW}
from the viewpoint of boundary state.
Let us consider a gauge theory on noncommutative $\R^n$, which is defined by
the algebra
\eqn\xcom{
[x^i,x^j]=i\th^{ij}.
}
On this space, the open Wilson line $W(k)$ is defined by
\eqn\Wkmattr{
W(k)=\Tr\exp\lf(-ik_i(x^i+\th^{ij}\h{A}_j)\ri)_{\st}
}
where $\h{A}$ is the noncommutative gauge field and the product of fields
is taken by the star product 
\eqn\defstar{
f\st g=f\exp\lf({i\o2}\overleftarrow{\partial_k}\th^{kl}
\overrightarrow{\partial_l}\ri)g.
}
In the boundary state formalism, $W(k)$ can be written as
an overlap between the closed string tachyon $|k\ket$ and the
boundary state $|B\ket$ describing a
$D$-brane in a constant $B$-field background \Ishibash: 
\eqn\WkinBoundary{
W(k)=\bra k|B\ket=\int[dx]\exp\lf(i\int_0^1d\si \hf x^iB_{ij}\del_{\si}x^j
-k_i(x^i+\th^{ij}\h{A}_j)\ri).
}
$\th$ and $B$ are related by
\eqn\Bvsth{
\th={1\o B}.
}
The closed string configuration $x(\si)$ is parametrized by $\si\in[0,1]$.

We can consider the general Wilson line which is the overlap between 
$|B\ket$ and the closed string momentum eigenstate:
\eqn\defgeneralWk{\eqalign{
\bra P(\si)|B\ket&=\Tr \,\P\exp\lf(-i\int_0^1 d\si P_i(\si)
(x^i+\th^{ij}\h{A}_j)\ri)_{\st} \cr
&=\int[dx]\exp\lf(i\int_0^1d\si \hf x^iB_{ij}\del_{\si}x^j
-P_i(x^i+\th^{ij}\h{A}_j)\ri),
}}
where $\P$ denotes the path ordering.

At first sight, it seems impossible to obtain the open contour 
from the boundary state which is an element of the Hilbert space
of a closed string. 
As shown in \refs{\IIKK,\DharW}, the open contour
does appear when we write the exponential in \Wkmattr\ as 
a product of small segments and move the factor $e^{-ikx}$ to the leftmost.  
In the rest of this section, we shall rederive this result from the path 
integral formalism. To do this, we should carefully treat the boundary 
condition of the integration variable $x(\si)$.
First we write $\bra P(\si)|B\ket$  as
\eqn\WPsiinperiodic{\eqalign{
\bra P(\si)|B\ket
&=\int dx_0 
\bra x_0|\P e^{-i\int_0^1 d\si P(\si)(x+\th\h{A})}|x_0\ket\cr
&=\int dx_0\bra x_0+y(1)|e^{-i\Pi(1)y(1)}\P
e^{-i\int_0^1 d\si P(\si)(x+\th\h{A})}
e^{i\Pi(0)y(0)}|x_0+y(0)\ket
}}
where $\Pi$ is the conjugate momentum of $x$. Here we introduced an arbitrary
path $y(\si)$ which may or may not be open at this stage. 
We can rewrite \WPsiinperiodic\ as a phase space 
path integral with constraints:
\eqn\pathintPix{\eqalign{
\bra P(\si)|B\ket=\int &[d\Pi dx_{{\rm open}}]
\cob(\chi_i)\det{}^{\hf}\{\chi_i,\chi_j\} \cr
&\cdot \exp\lf(i\int_0^1d\si 
\Pi_i\del_{\si}x^i_{{\rm open}}-P_i(x^i_{{\rm open}}+\th^{ij}\h{A}_j)
-\del_{\si}(\Pi_i y^i)\ri).
}}
The constraints are given by
\eqn\phii{
\chi_i=\Pi_i+\hf B_{ij}x_{\rm open}^j,
}
which are the second class: $\{\chi_i,\chi_j\}=B_{ij}$
\foot{For the BRST quantization of this system and its relation to
the Seiberg-Witten map, see \Rudychev.}.
The boundary condition of $x_{{\rm open}}$ is specified by 
\eqn\BCx{
x_{\rm open}(1)-x_{\rm open}(0)=y(1)-y(0).
}
Integrating out $\Pi$, $\bra P(\si)|B\ket$ becomes
\eqn\WIlsoninoen{\eqalign{
\bra P(\si)|B\ket=\int[dx_{\rm open}]\exp &\lf(i\int_0^1d\si  
\hf x_{\rm open}^iB_{ij}\del_{\si}x_{{\rm open}}^j
-P_i(x^i_{\rm open}+\th^{ij}\h{A}_j)\ri. \cr
&\qquad+\lf.{i\o2}\del_{\si}(y^iB_{ij}x_{\rm open}^j)\ri).
}}
In this expression, the contour $y(\si)$ is arbitrary. In the case
$\del_{\si}y=\th P$,  by shifting the variable 
$x_{\rm open}$ to $x+y$, $\bra P(\si)|B\ket$ can be written as
\eqn\periodicshift{\eqalign{
\bra P(\si)|B\ket&=\int_{\rm periodic}[dx]\exp \lf(i\int_0^1d\si  
\hf x^iB_{ij}\del_{\si}x^j-\hf y^iB_{ij}\del_{\si}y^j\ri.\cr
&\hskip 40mm\lf.+\del_{\si}(y^iB_{ij} x^j)
+\del_{\si}y^i\h{A}_i(x+y)\ri) \cr
&=e^{-i\int_0^1\hf y^iB_{ij}\del_{\si}y^j}
\Tr \,e^{-ipx}\P\exp\lf(i\int_0^1d\si\del_{\si}y^i\h{A}_i(x+y)\ri).
}}
In the last step, we used the relation
\eqn\yatBdry{
y(1)-y(0)=\int_0^1\del_{\si}y=\th p,
}
where $p$ is the center of mass momentum of closed string
\eqn\defsmallp{
p=\int_0^1d\si P(\si).
}

In summary, the open contour appears in the $y$-space which is defined by
\eqn\ytoP{
\del_{\si}y^i(\si)=\th^{ij}P_j(\si). 
} 
This is the same relation as the T-duality except for the factor of $\th$.
Therefore, we can understand the appearance of open contour as the 
winding mode which is T-dual to the center of mass momentum.
Note that the straight open line corresponds to the momentum $P(\si)$ without
non-zero modes, or the closed string tachyon $|k\ket$ (see \OkawaO\ 
for the recent discussion).

\newsec{Generalized Star Product}
In \refs{\startrekII,\MehenW}, it was shown that
the generalized star products appears when we expand the open Wilson lines.
In the path integral representation, the generalized star product 
can be written as a correlation function
of exponential operators on $\S^1$:
\eqn\defJascor{
\cob(k-\sum_ak_a)J_n(k_1,\ldots,k_n)=\lf\bra e^{-i\int d\si kx(\si)}
\prod_{a=1}^n\int_0^1 d\si_ae^{ik_ax(\si_a)}\ri\ket.
}
Here, the expectation value is defined by
\eqn\defbraket{
\Big\bra\O\Big\ket=\int_{{\rm periodic}}[dx]\O\exp\lf(i\int_0^1d\si
\hf x^iB_{ij}\del_{\si}x^j\ri).
}
The factor $\cob(k-\sum_ak_a)$ in \defJascor\
comes from the integral over the zero-mode 
of $x(\si)$. The non-zero mode $\til{x}(\si)$ of $x(\si)$ is defined by
\eqn\deftilx{
x(\si)=x_0+\til{x}(\si),\qquad \int_0^1d\si\til{x}(\si)=0.
}
The propagator of non-zero mode of $x$ on $\S^1$ 
should satisfy
\eqn\defGonS{
-iB_{ik}\del_1\bra\til{x}^k(\si_1)\til{x}^j(\si_2)\ket=\cob_i^j
\Big[\cob(\si_1,\si_2)-1\Big].
}
The solution of this condition is
\eqn\Gandtau{
\bra\til{x}(\si_1)\til{x}(\si_2)\ket=-{i\o2}\th\,\tau(\si_{12})
}
where $\si_{12}=\si_1-\si_2$ and
\eqn\deftau{
\tau(\si)=2\si-\ep(\si).
}
\fig{Graph of $\tau(\si)$ 
}{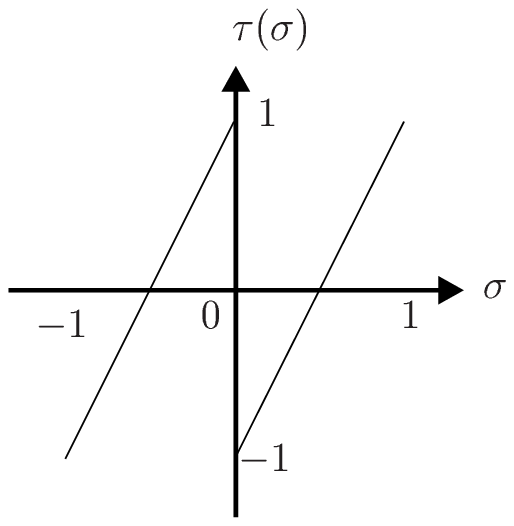}{50 truemm}
\figlabel\fragedtau
$\tau(\si)$ in \deftau\ is defined on the region $-1\leq\si\leq 1$.
We can extend $\tau(\si)$ to the outside of this region
with periodicity $1$. As a Fourier series, $\tau(\si)$ can be written as
\eqn\periodtau{
\tau(\si)=-{2\o\pi}\sum_{n=1}^{\infty}{1\o n}\sin(2\pi n\si).
}
$J_n$ in \defJascor\ is the correlation function of non-zero modes,
which is written as \startrekII
\eqn\Jnintau{
J_n(k_1,\ldots,k_n)=\prod_{a=1}^n\int_0^1d\si_a
\exp\lf({i\o2}\sum_{a<b}k_a\th k_b\tau(\si_{ab})\ri).
}
We denote the product determined by $J_n$ as $(f_1,\cdots,f_n)_n$.
Note that this product is totally symmetric.

By inserting
\eqn\insertdel{
1=\prod_{a=1}^{n-1}\int_{-1}^1ds_{an}\cob(s_{an}-\tau(\si_{an}))
}
and integrating over $\si_a\,(a=1,\cdots,n-1)$, $J_n$ can also be written as
\eqn\epintJn{
J_n={1\o 2^{n-1}}\prod_{a=1}^{n-1}\int_{-1}^1ds_{an}
\exp\lf({i\o2}\sum_{a<b}k_a\th k_bs_{ab}\ri)
}
where $s_{ab}\,(a,b\not=n)$ is given by
\eqn\tauijnonn{
s_{ab}=\tau\lf({s_{an}-s_{bn}\o2}\ri).
}
$J_n$ is known to satisfy the following descent relation \startrekII
\eqn\descent{
\hf\sum_{a=1}^nk_a\th k_nJ_n(k_1,\cdots,k_n)=\sum_{a=1}^{n-1}
\sin\lf(\hf k_a\th k_n\ri)J_{n-1}(k_1,\cdots,k_a+k_n,\cdots,k_{n-1}).
}
In the position space, this relation can be written as ({\it e.g.}
for $n=2,3$)
\eqn\descentot{\eqalign{
i\th^{kl}(\del_kf,\del_lg)&=[f,g]_{\st}, \cr
i\th^{kl}\del_k(f,g,\del_lh)_3&=(f,[g,h]_{\st})+(g,[f,h]_{\st}),
}}
where $[f,g]_{\st}=f\st g-g\st f$ and $(f,g)\equiv (f,g)_2$.

\newsec{Seiberg-Witten Map and Generalized Star Product}
In this section, we consider the Seiberg-Witten map 
from the viewpoint of the  gauge equivalence condition \SW\ and
the two descriptions of an open Wilson line \refs{\Ishibash,\Okuyama}. 

\subsec{Seiberg-Witten Map from Gauge Equivalence Condition}
First we consider the Seiberg-Witten map as the correspondence
between commutative and noncommutative gauge transformation.
To construct the map, it is more convenient to see this map as the 
correspondence of two BRST transformations.
The nilpotent BRST transformation 
associated with commutative gauge transformation is
\eqn\BRSTcom{
\cob A_i=\del_ic,\quad \cob c=0.
}
For the noncommutative case, the BRST transformation is given by   
\eqn\BRSTNcom{
\cob \h{A}_i=\del_i\h{c}-i[\h{A}_i,\h{c}\,]_{\st},\quad
\cob \h{c}={i\o2}[\h{c},\h{c}\,]_{\st}.
}
$c$ and $\h{c}$ are ghost fields for commutative and noncommutative
gauge symmetries, respectively. Here, the commutator of the general fields
$f,g$ with ghost number $|f|,|g|$ 
is defined by $[f,g]_{\st}=f\st g-(-1)^{|f||g|}g\st f$.
Note that $[\h{c},\h{c}\,]_{\st}$ does not 
vanish since $\h{c}$ carries the ghost number $1$.

We can construct the Seiberg-Witten map as a power series in $A$:
\eqn\mapAc{\eqalign{
\h{A}&=A+\sum_{n=2}^{\infty}A^{(n)}=\h{A}(A), \cr
\h{c}&=c+\sum_{n=1}^{\infty}c^{(n)}=\h{c}\,(c,A),
}}
where the superscript $n$ denotes the order of $A$.
First, let us consider the map for ghosts. 
By equating the terms of the same order in $A$, the relation 
$\cob\h{c}=i[\h{c},\h{c}\,]_{\st}/2$ can be decomposed as 
\eqn\eqforc{\eqalign{
\cob c^{(1)}&={i\o2}[c,c]_{\st}=-\hf\th^{kl}(\del_kc,\del_lc), \cr
\cob c^{(2)}&=i[c,c^{(1)}]_{\st}.
}}
Here, we used the relation \descentot.
By ``integrating'' this relation, we find that $c^{(1)}$ and $c^{(2)}$
are given by 
\eqn\solforc{\eqalign{
c^{(1)}&=-\hf\th^{kl}(A_k,\del_lc), \cr
c^{(2)}&=-\hf\th^{kl}\th^{mn}\bra A_k,\del_mc,\del_lA_n\ket,
}}
where $\bra \,\ket$ is defined by
\eqn\defbraket{
\bra f,g,h\ket=(f,(g,h))+(g,(f,h))-(f,g,h)_3.
} 
This product is symmetric in the first two arguments:
$\bra f,g,h\ket=\bra g,f,h\ket$. 
Now we consider the map for gauge fields. 
The equations satisfied by $A^{(2)}$ and $A^{(3)}$ are
\eqn\eqforAn{\eqalign{
\cob A_i^{(2)}&=\del_ic^{(1)}-i[A_i,c\,]_{\st}, \cr
\cob A_i^{(3)}&=\del_ic^{(2)}-i[A_i^{(2)},c\,]_{\st}-i[A_i,c^{(1)}]_{\st}. 
}}
The solution for these conditions is found to be
\eqn\solforA{\eqalign{
A^{(2)}_i&=-\hf\th^{kl}(A_k,\del_lA_i+F_{li}), \cr
A^{(3)}_i&=-\hf\th^{kl}\th^{mn}\Big[\bra A_k,\del_mA_i,\del_lA_n\ket
+\bra A_k,F_{mi},F_{ln}\ket
+\bra A_k,A_m,\del_lF_{in}\ket\Big].
}}

\subsec{Path Integral Derivation of Seiberg-Witten Map}
In this subsection, we derive the Seiberg-Witten map from the 
open Wilson line in commutative and noncommutative pictures.
We will see that the map obtained by this method agrees
with that in the previous subsection.
The relation between $A$ and $\h{A}$ can be written as
\eqn\SWmapWhW{
\h{W}(\h{A})=W(A)
}
where
\eqn\SWincorr{\eqalign{
\h{W}(\h{A})&=\lf\bra\exp\lf(-ipx(0)+i\int_0^1 d\si 
\h{A}_i(x+y)\del_{\si}y^i\ri)\ri\ket, \cr
W(A)&=\lf\bra\exp\lf(-ipx(0)+i\int_0^1 d\si 
A_i(x+y)\del_{\si}(x^i+y^i)\ri)\ri\ket.
}}

As discussed in \Okuyama, Seiberg-Witten map can be extracted from 
\SWmapWhW\ by expanding the exponential  and equating the term 
with single $\si$-integral. 
The term with single $\si$-integral in $\h{W}(\h{A})$ is 
\eqn\hWwithsi{
i\int_0^1 d\si \del_{\si}y^i\h{A}_i(p)e^{ipy(\si)}.
} 
The corresponding term in $W(A)$ has the form
\eqn\WAexp{
i\int_0^1d\si \del_{\si}y^i\sum_{n=1}^{\infty}A_i^{(n)}(p)e^{ipy(\si)}
}
where $A^{(n)}$ is the $n$-th order term of $A$.
Let us compute the first three terms of this expansion.
As easily seen, $A^{(1)}$ is equal to $A$. $A^{(2)}$ is found to be
\eqn\defAtwo{
A_i^{(2)}(p)=-\hf\th^{kl}\int dk_1dk_2\cob(p-k_1-k_2)I_2(k_1,k_2)
A_k(k_1)(\del_lA_i+F_{li})(k_2)
}
where $I_2$ is given by
\eqn\defItwo{
I_2(k_1,k_2)=\int_0^1d\si_1\cob(\si_{12})
\exp\lf({i\o2}k_1\th k_2\tau(\si_{12})\ri).
}
The factor $\cob(\si_{12})$ 
comes from the contraction between $\del_1x(\si_1)$ 
and $x(\si_2)$.
By regularizing the $\cob$-function as in \Okuyama, we find 
that $I_2$ is equal to $J_2$:
\eqn\ItwoisJtwo{
I_2(k_1,k_2)=\hf\int_{-1}^1d\ep_{12}\exp\lf(-{i\o2}k_1\th k_2\ep_{12}\ri)
=J_2(k_1,k_2).
}
Therefore, \defAtwo\ reproduces the result \solforA\ obtained from the 
gauge equivalence relation. 
 
Let us consider $A^{(3)}$.
From \SWincorr, $A^{(3)}$ is found to be
\eqn\AthtoIth{\eqalign{
A^{(3)}_i(p)=-&\hf\th^{kl}\th^{mn}
\int\prod_{a=1}^3dk_a\cob(p-\sum_ak_a)I_3(k_1,k_2,k_3) \cr
&\cdot A_k(k_1)\Big[\del_mA_i(k_2)\del_lA_n(k_3)+F_{mi}(k_2)F_{ln}(k_3)
+A_m(k_2)\del_lF_{in}(k_3)\Big]
}}
where $I_3$ is given by
\eqn\Ithdef{\eqalign{
I_3&=\int_0^1d\si_1d\si_2\cob(\si_{13})\cob(\si_{23})
\exp\lf({i\o2}\sum_{a<b}^3k_a\th k_b\tau(\si_{ab})\ri) \cr
&=\prod_{a=1}^3\int_0^1d\si_a\cob(\si_{13})\cob(\si_{23})
\exp\lf({i\o2}\sum_{a<b}^3k_a\th k_b\tau(\si_{ab})\ri).
}}
Here we put an extra integral over $\si_3$ since $I_3$ does not depend on 
$\si_3$.
To evaluate $I_3$, the following identity is useful: 
\eqn\identity{
{1\o4}\prod_{a=1}^3\int_0^1d\si_a\del_1\tau(\si_{13})\del_2\tau(\si_{23})
\exp\lf({i\o2}\sum_{a<b}k_a\th k_b\tau(\si_{ab})\ri)=0.
}
This identity holds due to the periodicity of $\tau(\si)$.
Using the relation $\hf\del_{\si}\tau(\si)=1-\cob(\si)$, 
$I_3$ can be written as
\eqn\Ithinmoon{\eqalign{
I_3&=\prod_{a=1}^3\int_0^1d\si_a\Big[\cob(\si_{13})\cob(\si_{23})
-(1-\cob(\si_{13}))(1-\cob(\si_{23}))\Big]
\exp\lf({i\o2}\sum_{a<b}k_a\th k_b\tau(\si_{ab})\ri) \cr
&=\prod_{a=1}^3\int_0^1d\si_a\Big[\cob(\si_{13})+\cob(\si_{23})-1\Big]
\exp\lf({i\o2}\sum_{a<b}k_a\th k_b\tau(\si_{ab})\ri) \cr
&=J_2(k_1,k_3)J_2(k_1+k_3,k_2)+J_2(k_2,k_3)J_2(k_1,k_2+k_3)-J_3(k_1,k_2,k_3).
}}
We can see that the product defined by 
$I_3$ is nothing but $\bra f,g,h\ket$ in \defbraket.
Hence \AthtoIth\ agrees with \solforA.

Although we cannot write down the tensor structure of spacetime indices $i,j$ 
for the general term $A^{(n)}$,
we can determine the form of product appearing in $A^{(n)}$.
$A^{(n)}$ is associated with the integral with $n-1$ $\cob$-functions.
This set of $\cob$-functions determines the tree graph with $n$ vertices
and $n-1$ links. All the vertices should be connected by this graph
in order for $A^{(n)}$ to be a single $\si$-integral term. 
The integral associated with the graph $G$ is given by
\eqn\InG{\eqalign{
I_n(G)=&\prod_{a=1}^n\int_0^1d\si_a\lf[\prod_{L:{\rm link}}
\cob(\si_{L_iL_f})-\prod_{L:{\rm link}}(1-\cob(\si_{L_iL_f}))\ri]
\exp\lf({i\o2}\sum_{a<b}k_a\th k_b\tau(\si_{ab})\ri) \cr
=&-J_n(k_1,\cdots,k_n)
+\sum_{L:{\rm link}}J_2(k_{L_i},k_{L_f})
J_{n-1}(k_1,\cdots,k_{L_i}+k_{L_f},\cdots,k_n) +\cdots,
}}
where $L_i,L_f\in \{1,\cdots,n\}$ are the end points of $L$.

\subsec{Area Derivative of Open Wilson Line}
Using the boundary state picture of open Wilson lines, 
we can easily construct the field strength operator 
with open Wilson line attached. This operator was used in \MehenW\
to construct the Seiberg-Witten map in the form $A=A(\h{A})$.
The open Wilson line in the commutative and noncommutative picture 
is written as
\eqn\braPketB{\eqalign{
\bra P(\si)|B\ket&=\int[dx]\exp\lf(i\int_0^1d\si \hf x^iB_{ij}\del_{\si}x^j
-P_i(\si)\phi^i(x(\si))\ri) \cr
&=\int[dx]\exp\lf(i\int_0^1d\si \hf x^iB_{ij}\del_{\si}x^j+A_i(x)\del_{\si}x^i
-P_i(\si)x^i(\si)\ri), 
}}
where 
\eqn\xtoX{
\phi^i=x^i+\th^{ij}\h{A}_j.
}

By performing the ``area derivative'' of open Wilson line defined by
\eqn\defarea{
{\cob\o\cob\Si^{ij}}=\lim_{\ep\riya+0}{\cob\o\cob P^i(\ep)}{\cob\o\cob P^j(0)}
-{\cob\o\cob P^i(0)}{\cob\o\cob P^j(\ep)},
}
we find that
\eqn\eqXXtoxx{
\Tr_{B}\Big([\phi^i,\phi^i]e^{-ik\phi}\Big)_{\st_{B}}
=\Tr_{{\cal F}}\Big([x^i,x^i]e^{-ikx}\Big)_{\st_{{\cal F}}}.
}
Here $\st_B$ and $\Tr_B$ are the star product and the trace defined by
the symplectic form $B$. 
In the commutative side, the symplectic form becomes
\eqn\BtoF{
{\cal F}=B+F=\th^{-1}+F.
}
As discussed in \startrekII, this relation \eqXXtoxx\ 
can be understood as the equivalence of
star products:
\eqn\equivalence{
Tf\st_{B}Tg=T(f\st_{{\cal F}}g).
}
With this operation $T$, $\phi$ and $x$ are related by $\phi^i=T(x^i)$. 
See \JurcoS\ for the derivation of the Seiberg-Witten map from the
equivalence of star products.

The left-hand-side of \eqXXtoxx\ can be evaluated as
\eqn\NChF{
\Tr_{B}\Big([\phi^i,\phi^i]e^{-ik\phi}\Big)_{\st_{B}}
=i\Tr_B\lf[(\th-\th\h{F}\th)^{ij}\P\exp\lf(i\int_0^1d\si 
\h{A}_i(x+\th k\si)\th^{ij} k_j\ri)e^{-ikx}\ri]_{\st_B}.
}
The right-hand-side of \eqXXtoxx\ is a complicated function 
of $F$, which can be calculated in principle by the formula 
in \refs{\Kont,\CF}. If we neglect the derivative of $F$, the commutator of
$x$ can be written as
\eqn\xcom{
[x^i,x^j]_{\st_{{\cal F}}}=i({\cal F}^{-1})^{ij}+\O(\del F)
=i\lf({1\o\th^{-1}+F}\ri)^{ij}+\O(\del F).
}

\newsec{Conclusion and Discussion}
In this paper, we considered the open Wilson line as a momentum 
representation of the boundary state in a constant $B$-field background.
The straight open Wilson line corresponds to the tachyonic particle
state of closed string, and the curved open Wilson line 
appears as the coupling between a $D$-brane and an extended closed string.
The generalized star products can be understood as the correlation functions
of exponential operators on the boundary of worldsheet. The Seiberg-Witten map
can be obtained by comparing the open Wilson line written in commutative and 
noncommutative pictures.
The products of fields appearing in the Seiberg-Witten map are 
the combination of the generalized star products, and in principle 
the general term of the map can  be determined by the rule of Wick contraction.

In \startrekII, a closed form for the expression of
the ordinary field strength in terms of the noncommutative 
gauge field was proposed.
Although \eqXXtoxx\ gives the exact relation between 
$F$ and $\h{A}$, it seems difficult to prove eq.(1.7) in \startrekII.

In \OkawaO, it was shown that the interactions 
between the noncommutative gauge field and closed string modes 
in the superstring theory 
are different from those in the bosonic string 
theory. It is interesting to 
extend our result to the case of superstring.

\vskip 12mm
\centerline{{\bf Acknowledgments}}
I would like to thank N. Ishibashi  
for useful comments and discussions. 
This work was supported in part by JSPS Research Fellowships for Young
Scientists.

\listrefs

\end